# Monitoring, assessment and numerical analysis of the masonry bridge pier defects

**Pavel Ryjáček,**

*Faculty of Civil Engineering CTU in Prague, Prague, Czech Republic*

**Christos Mourlas**

*University of Dundee, Dundee, United Kingdom*

**David Citek**

*Klokner Institute, CTU in Prague, Prague, Czech Republic*

**Contact:** pavel.ryjacek@fsv.cvut.cz

## Abstract

The paper thoroughly analyses the significant cracking on the newly refurbished masonry bridge pier in Lovosice. The cracks of width up to 15 mm appeared shortly after the reconstruction of the bridge and were the subject of a detailed assessment. First, a monitoring system was mounted to analyse the behaviour during the year and the dependency on the temperature. Next, a dive inspection was done in order to evaluate the conditions under the water. The next step was to make a diagnostics survey, based on the drilling, endoscopic tests and water pressure tests. Based on the results, several detailed numerical models were created in order to analyse the pier behaviour and conduct a parametric study to investigate various factors, leading to the defects. The numerical models covered the complex geometry of the pier, consisting of the inner poor concrete core and outer shell from sandstone masonry blocks.  The above-described procedure revealed the causes of the defects and also recommended the proper strengthening of the pier.

## 1   Introduction

The bridge in Lovosice, Czech Republic, was recently fully refurbished. The superstructure was removed and replaced by a new one, but the substructure remained and was partially strengthened. Unfortunately, shortly after the reconstruction finished, cracks of width up to 15 mm appeared in Pier P3 and developed fast. Because of the fears of the railway administration, a detailed analysis was conducted to identify the causes of the cracks. The scope of the work included the following:

- Detailed diagnostics and underwater inspection.
- Development of a 3D numerical model and analysis of the pier's behavior.
- Validation of the numerical results with the monitoring data.
- Determination of the causes of damages and suggested solutions.

The combination of sensing technology for monitoring and the development of 3D finite element modeling is used to tackle the challenging task of capturing the behavior of masonry structures. A state-of the art literature review can be found in [1].

## 2  Bridge description

The load-bearing structure is designed as a welded steel structure with rigid beam main girders, braced by arches with vertical hangers (so-called Langer beam). The structure is designed with a lower orthotropic bridge deck. From the structural point of view, it consists of three simple spans with a span of 74,37 m.

The main girder is a welded I-beam of 2,5 m height, with an arch of closed cross-section. The span of the arch is approximately 74,3 m. The vertical hangers were designed from solid round bars Ø110 mm welded into the contact plates. The distance between the main beams is 6.050 m. The bridge deck is designed as an orthotropic structure. The longitudinal ribs are carried at an axial distance of 460 mm, while the transverse stiffeners follow a regular modulus of 2,010 m. A view of the bridge can be seen in Figure 1.

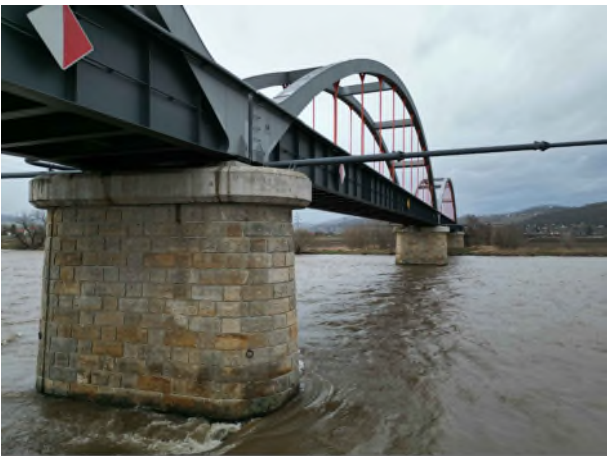

*Figure 1. The view on the bridge*

The pier was constructed with an external layer of masonry sandstones connected with mortar and it is filled with concrete and more masonry stones. Because of the good condition of the piers and the aesthetical quality, connected with the heritage protection principles, it was decided to keep the piers also for the new superstructure. The masonry stones inside the pier form a brick cup-layer that encloses the filled concrete at different levels of height. In this way, the pier is consisted of 5 brick boxes filled with concrete as can be seen in Figure 2. The masonry region has 9.3 m height, 10.5 m width and 4.224 m thickness as can be seen in Figure 2 and Figure 3.

The cross-section of the pier consists of an external brick layer having an ellipsoid shape which is filled with concrete. Every 1.5m height of the pier, the cross section is filled with bricks to form a brick cup. The cross-section can be shown in Figure 2.

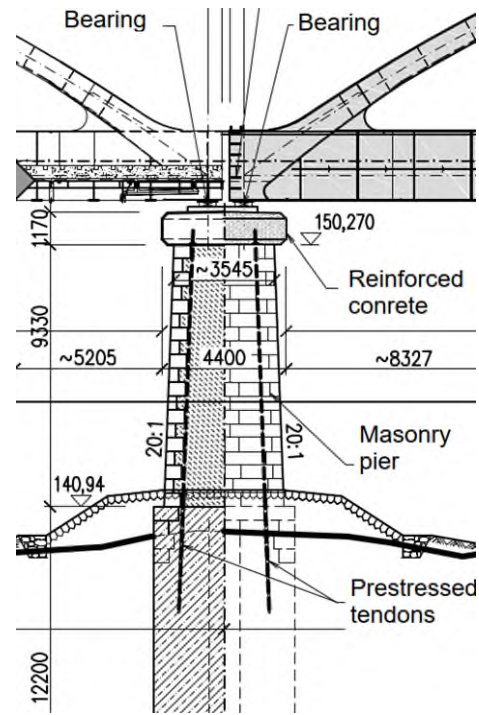

*Figure 2. The cross section of the pier*

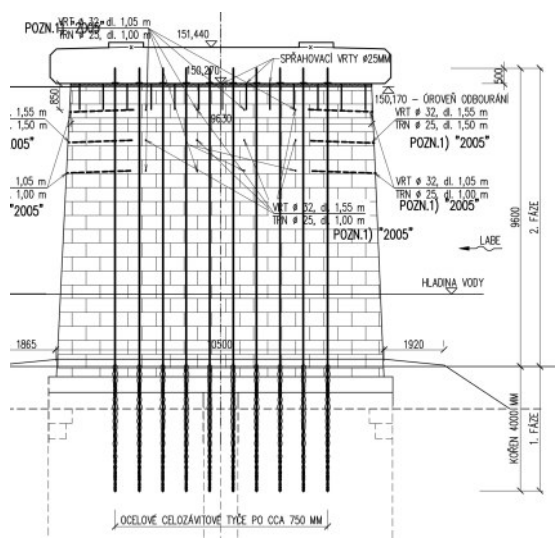

*Figure 3. The side view on the pier*

In the top of the pier, there is a reinforced concrete plate where the bearings are placed. The plate appears to be heavily reinforced. The foundation of the pier has been made with the 'caisson' method.

In the top of the masonry pier region below the concrete plate, there are some tendons both horizontal (with a small inclination) and vertical of length that vary from 1 to 1.5 m. In addition to the previous tendons, there are 10 prestressed tendons oriented vertically to the pier's cross-section (as shown in Figure 2 and Figure 3), each measuring 13 m in length with a prestressing force of 100 kN.

## 3 Diagnostics of the defects

### 3.1 Visual inspection

In order to estimate the dimensions of the stones and the thickness of the mortar, a drone was used to take several pictures of the pier, they were used for the 3D model creation. The masonry structure consists of big and small blocks of the same height.

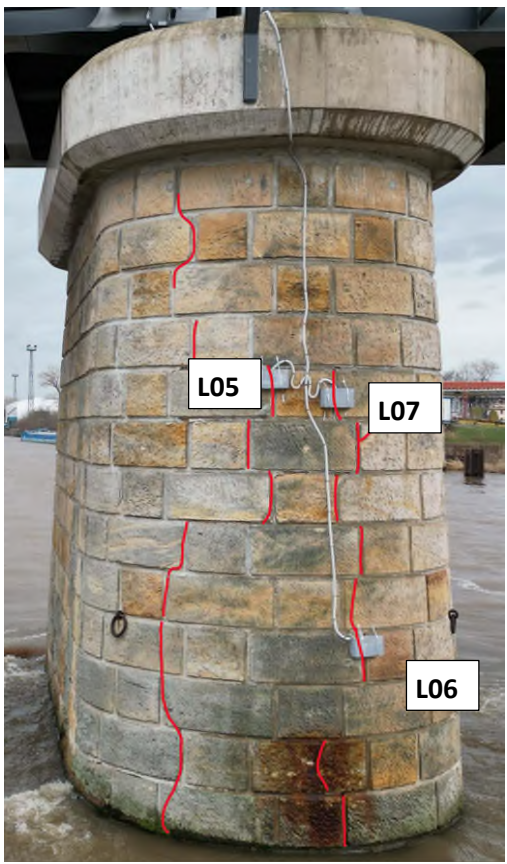

Figure 4. The pier P3, red lines show the location of the cracks and the labels of the sensors

An inspection has been conducted including diving into the river to detect the length of the cracks of the examined pier, as shown in Figure 4. The inspection shows that the cracks below the water are starting to fade away after a certain depth and at the river bottom to disappear. It means, that the cause is not due to the insufficient capacity of the foundation.

### 3.2 Drilling inspection and water pressure tests

Based on the visual findings, drilling holes were made, and the inner structure of the pier was examined. Three holes were made on the front side, where the crack size was the largest, as shown in Figure 5.

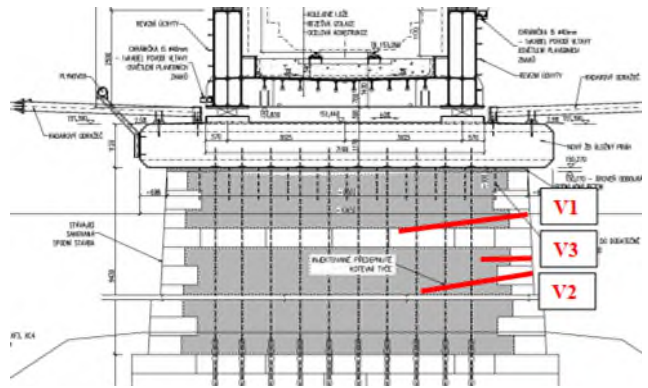

Figure 5. The pier P3, red lines show the location of drills

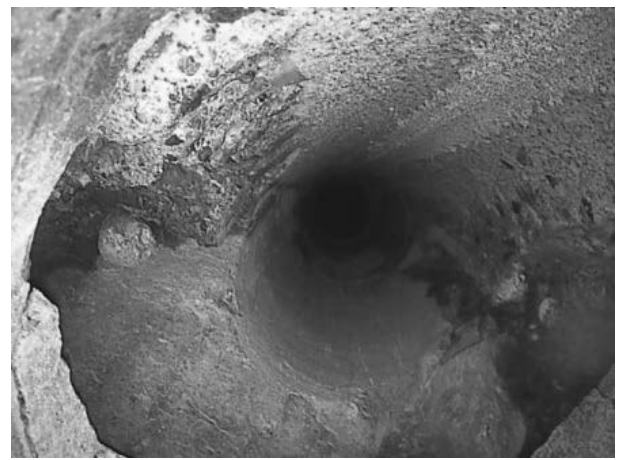

Figure 6. The pier P3, endoscopic image in the drill, large caverns are visible

The water pressure tests showed, that the water was freely leaking through the inner core (as shown in Figure 6), it was not possible to reach any pressure. That means, that 10% porosity can be expected.

The diagnostic also included the analysis of the stones used in the construction of the pier and tesring of their compressive strength. The characteristic strength of the outer masonry was 7.9 MPa and the inner masonry was 5.8 MPa.

## 4 Long-term monitoring

The LVDT displacement sensors (called also crackmeters) were located on the front side of the pier, as shown in Figure 4. As shown in Figure 7, the development of the crack depends strongly on the temperature (as shown in Figure 8), and in a short-time time, the relation is linear.

The monitoring was provided by the Pontex Ltd. engineering company.

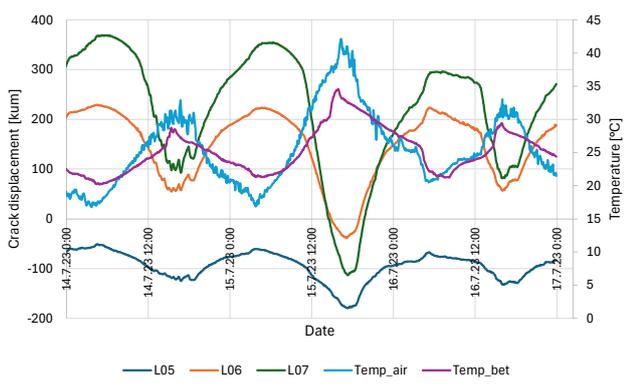

*Figure 7. The pier P3, endoscopic image in the frill, large caverns are visible*

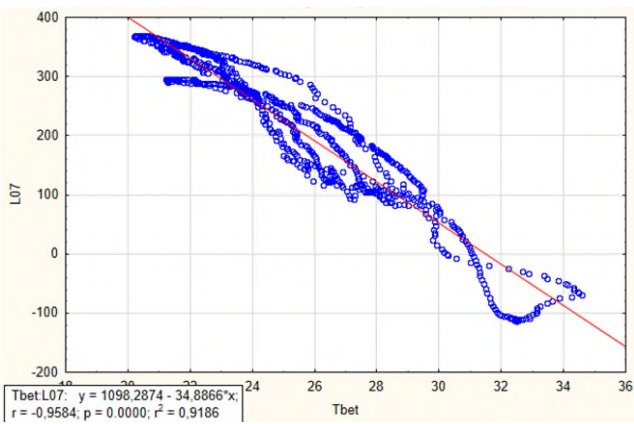

*Figure 8. The relation between opening of the crack in location L05 and inner temperature of the pier $T_{bet}$*

However, when considering long-term behavior, the relationship remains linear, but the presence of hysteresis loops (as shown in Figure 9) suggests that this could be due to the non-linear behavior of the masonry material. It may also indicate the influence of forces from the rail-track interaction, which follow the same hysteresis principle.

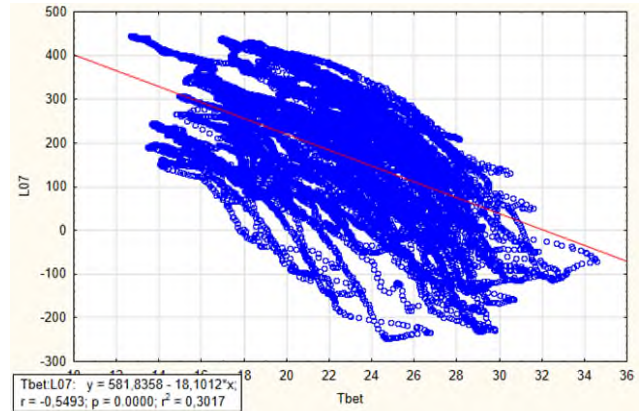

*Figure 9. The relation between displacement in L07 and temperature in the total t ime-period*

## 5 Numerical modelling

The pier has been modelled with the use of Abaqus. The methodology of modelling the masonry structure follows Abdulla et al. [2] using a micro-modelling approach. The method is able to capture the crack propagation within the masonry blocks without predefining the possible crack locations in the model. The simplicity of the method lies in the fact that it is available in the Abaqus library without employing any user-defined subroutines.

The accurate modelling of masonry structures is an open subject among the scientific community. There are many papers that propose different approaches in order to handle the interaction problem between different materials and the contact properties. An exhausting literature review can be found in Altri et al. [3]. According to this study, the different approaches of modelling masonry structures can be classified as:

1. Block based models (interface element-based, contact-based, textured continuum, limit analysis and extended finite element approaches).
2. Continuum homogeneous models (direct and multiscale approaches using homogenization procedures)

3. Geometry-based models (rigid bodies limit state equilibrium based on static and kinematic theorems)
4. Macroelement models (equivalent beam-based and spring-based approaches)

As concerned the finite element approaches used for masonry structures, the approaches can be simply divided in three categories: i) detailed micro-model [4,5], ii) simplified micro-model [6,2] and iii) macro-model [7,8].

The detailed micro-modelling approach is a model that simulates both the mortar and the block-units as continuum elements. In contrast, with the simplified micro model, the block units are expanded by adding the mortar thickness. In this approach, the expanded blocks are modelled as continuum elements and the interaction between them as discontinuum elements. The last approach (macro-model) considers masonry as a homogenous material without distinguishing between mortar and block units.

In the present work, the simplified micro-modeling approach is employed, utilizing average material properties for the combined behavior of both materials. The discontinuity in the contact elements is modelled using the extended finite element method (XFEM). In this approach, a discontinuous enrichment function is added to the FEM formulation based on the partition of unity concept proposed by [9]. In this study, the XFEM cracks are modelled according to the segment method [10]. The model of half of the pier can be seen in Figure 10, highlighting the different parts with varying material characteristics used in the analyses. The model with the embedded tendons is shown in Figure 11.

The expanded block-units are modelled using 3D hexahedral eight-nodded solid elements in Abaqus/Standard software. For the constitutive model of the block-units, the Drucker-Prager plasticity model is used. The interfaces (mortar) are modelled through the contact friction-cohesive properties. The contacts of the adjacent masonry units are defined through the sliding properties. In addition, it is assumed that hard contact behavior is applied between the adjacent surfaces of the masonry units. This implies that the penetration of the masonry block-units is prevented throughout the analysis.

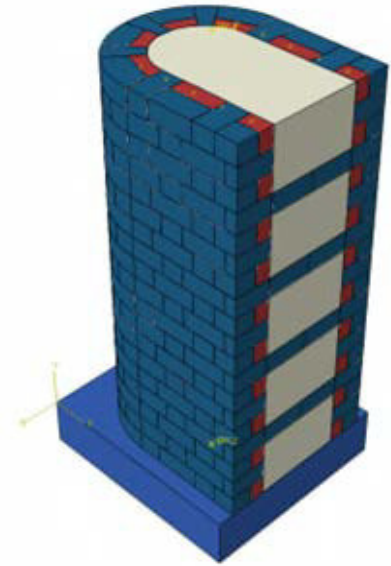

Figure 10. The view of the model, showing the main parts – blue – masonry blocks, white – inner concrete core, red – variable material blocs for cavern simulation

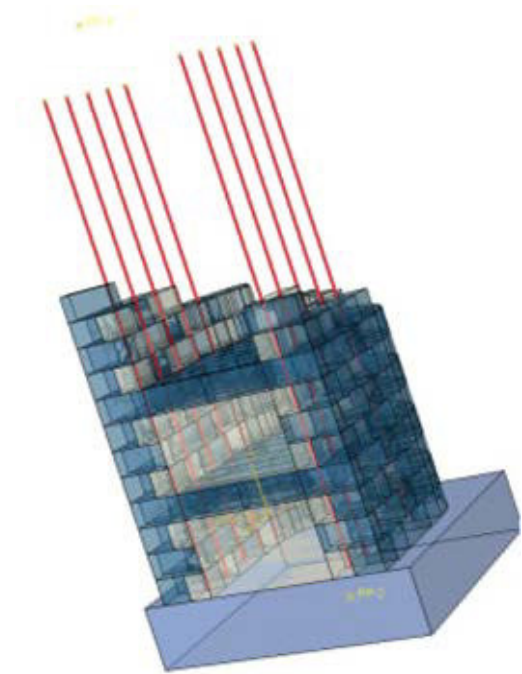

Figure 11. The view of the model, including the prestressing tendons.

The filled concrete is modelled with 8-nodded hexahedral elements too. Additionally, the steel reinforced concrete plate that exists at the top of the pier is also modelled by 8-noded hexahedral elements and it is considered a rigid body (a sparse meshing is used). The same modelling assumption is used for the foundation of the pier. The tendons are modelled as beam elements which are tied to the surrounding concrete as embedded elements. The half pier is analyzed under the half of the vertical and horizontal loading that corresponds to the ultimate state according to the design and the values are: $R_v$ = 15395/2=7697,5 KN and $R_h$ = 1958/2=979 KN respectively.

### 5.1 Damage scenarios

Several scenarios were investigated, in order to analyze the reasons for the crack's formation. They were the following:

- Deterioration of the material properties by removing modelling parts.
- Deterioration of the material properties, modelled by varying the material properties and modulus of elasticity (E-modulus).
- Deterioration of the strength properties of brick-units, modelled by decreasing the angle of friction.
- Deterioration of the interaction properties of brick-units, modeled by reducing the coefficient of friction.
- Deterioration of the material properties in certain areas in the concrete area (implying the presence of voids or holes).
- Deterioration of the material properties in certain areas (implying the presence of holes) in the concrete area, combined with a reduction in the coefficient of friction, implying the absence of mortar.

In general, achieving a similar level of defects was challenging. However, in the final case—where friction was reduced and voids were present—crack openings appeared in the front part of the pier, along with separations at the block interfaces. The distribution of the slippage between the brick-stones is depicted in Figure 12.

This investigation has been conducted assuming several material parameters as an attempt to estimate the vulnerability of this complicated system and explain the appearance of the cracks in the pier. The numerical model showed that all the components contribute significantly to the stiffness and strength of the pier. In Figure 13, the stress distribution of the filled parts (concrete and brick-stones) is illustrated.

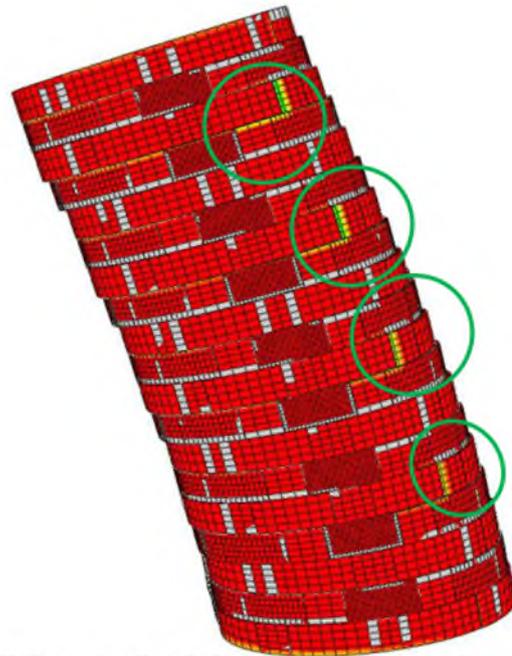

*Figure 12 Distribution of CSLIP1 variable for the last modelling approach*

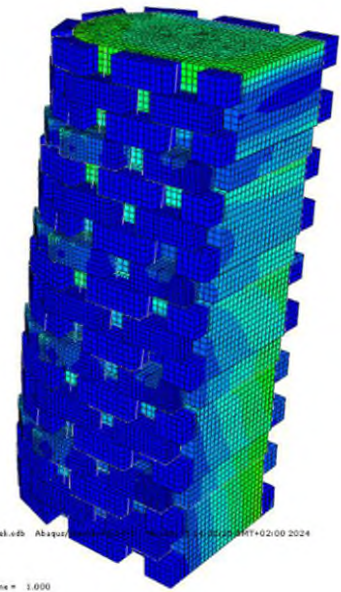

*Figure 13. The distribution of Von Mises stresses in the pier.*

Specifically, according to the numerical analysis, the properties of the filled concrete material and the brick material exhibit specific limitations. The compressive strength of the filled concrete should be approximately 28 MPa with a Young's Modulus of around 27 GPa. Regarding the brick stones, the Young's Modulus of should be around 1200 MPa.

As concerns the holes and the voids that exist inside the pier, it is evident that the magnitude of this damage should be thoroughly tested. According to the numerical analysis, the removal of elements that are the size of a brick stone inside the pier has destructive consequences for the designed limit state of the pier. If the size of the holes is estimated in a certain level of precision, then a new shear strength of the structure should be calculated.

If the damage is not excessive and does not compromise the global stability of the structure, it can be modeled by deteriorating the material properties. The numerical analysis shows that if the layer of concrete that is in contact with the external brick wall is deteriorated significantly (E=15GPa, fc=15MPa), then the analysis does not appear to have significant changes in terms of stresses. However, the deformability of the system is significantly increased. In addition to that, the connection properties significantly affect the deformability of the system. An estimation of the coefficient of friction lower than 0.6 can lead to significant slippage that can cause crack development. It is clear that the damage of the mortar should be investigated experimentally in order to simulate accurately the interface properties of the numerical model.

## 6  Findings and conclusions

According to the results, the crack behaviour seems to be influenced by a change in the temperature of the pier and the associated change in the width of the blocks, but probably also by the horizontal load from the continuous welded track (rail-track interaction). The crack opening was increasing, but it did not reach the critical phase.

Numerical analysis was performed assuming several material parameters in an attempt to explain the crack initiation in the pier. From a global point of view, the analysis shows that the pier behaves stably, with the stresses developed remaining globally below the ultimate strength.

The removal of elements within the pier - the cavern - has destructive consequences, creating locally overstressed spots and slips, which may manifest themselves precisely as cracks.

In addition, the properties of the joints significantly affect the deformability of the system. Friction coefficients below 0.6 can lead to significant slippage, which can also cause excessive cracks.

The extent of the reconstruction in 2005 is believed to have been suboptimal. Although limited grouting of the pier was reported, it is unlikely that these faults and caverns could have developed in just 10-15 years. Unfortunately, before the 2015 reconstruction, the pier was assumed to be repaired and in good condition, which was not the case. This misconception led the designer to conduct inadequate investigations based on the incorrect assumption.

Nevertheless, the project prescribed grouting and reinforcement with rods - given the extent of knowledge at the time, this seems a safe approach. When the gaps were identified during construction, the work carried out was appropriate to the defects identified and the installation of the planned rods involved grouting to a much larger cubic capacity than that specified in the design. Unfortunately, as per the design, the rods and the boreholes were not at the front part of the piers, which resulted in this area not being grouted.

This is obvious from today's perspective; however, the problem could have been foreseen at the time. It should have been addressed by increasing the extent of grouting and ensuring the quality of the grouting through pressure tests.

From the point of view of the cause of the cracks - the considerable caverns, insufficient grouting and missing mortar in the joints led to the degradation of the grouting. The existing caverns then caused local stresses, leading to the formation of cracks. It is also believed that horizontal forces from interaction with the continuous welded track may have had some influence.

It is believed from the current investigation that the presented cracks, was not a clear and unambiguous error by a single entity, but rather a chain of problems and deficiencies based on partial assumptions and hypotheses, particularly the assumption that the pier was adequately repaired in 2005. Other issues then compounded the situation.

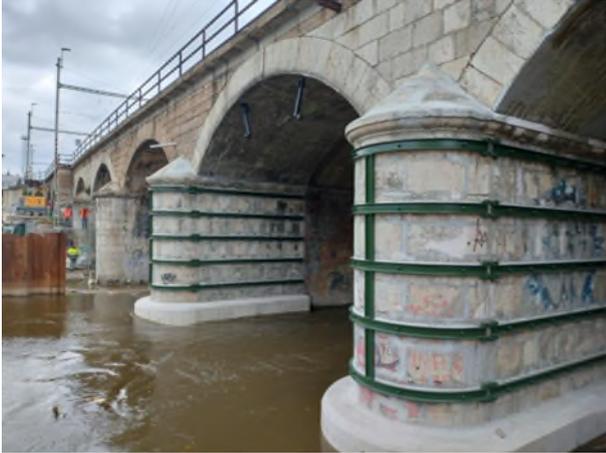

*Figure 14 Example of the pier, strengthened in Brno, Czechia*

As the final proposed solution to the problem, it is suggested to strengthen the pier, with grouting of the pier with microcement suspension and to prestress the pier with steel hoops, as shown in Figure 14.

## 7 Acknowledgements

*The paper is partly funded by the project No. TM05000045 Intelligent railway infrastructure monitoring system, funded by Technological agency of Czech Republic.  The paper is also partly in the implementation part funded by the project No. CZ.02.01.01/00/23_020/0008487 (INODIN) co-funded by European Union. For the purpose of Open Access, a CC BY 4.0 public copyright license has been applied by the authors to the present document.*

## 8 References

[1] Polepally, G., Pasupuleti, V.D.K. & Kalapatapu, P. A comprehensive survey of masonry bridge assessment methods: past to present. *Innov. Infrastruct. Solut.* **9**, 219 (2024).

[2] Abdulla KF, Cunningham LS, Gillie M (2017) Simulating masonry wall behaviour using a simplified micro-model approach. Eng Struct 151:349–365.

[3] D'Altri, A.M., Sarhosis, V., Milani, G. et al. Modeling Strategies for the Computational Analysis of Unreinforced Masonry Structures: Review and Classification. Arch Computat Methods Eng 27, 1153–1185 (2020).

[4] Greco, F., Leonetti, L., Luciano, R., Pascuzzo, A., Ronchei, C., A detailed micro-model for brick masonry structures based on a diffuse cohesive-frictional interface fracture approach, Procedia Structural Integrity,Volume 25,2020,Pages 334-347.

[5] Cross, T., De Luca, F., De Risi R., Camata, G., Petracca, M. Micro-modelling of stone masonry template buildings as a strategy for seismic risk assessment in developing countries, Engineering Structures,Volume 274,2023, 114910.

[6] Abasi A., Banting B., Sadhu A, Strength evaluation of early-age full-scale unreinforced masonry walls against out-of-plane loading using experimental and numerical studies, Eng. Struct., Volume 325,2025,119507

[7] Lourenço P B, Milani G, Tralli A and Zucchini A 2007 Analysis of masonry structures: review of and recent trends in homogenization techniques *Can. J. Civ. Eng.* **34** 1443–57.

[8] Asteris P G, Sarhosis V, Mohebkhah A, Plevris V, Papaloizou L, Komodromos P and Lemos J V 2015 Numerical modeling of historic masonry structures *Handbook of Research on Seismic Assessment and Rehabilitation of Historic Structures* ed P Asteris and V Plevris (IGI Global, Hershey, PA) 213–56.

[9] Melenk JM, Babuška I. The partition of unity finite element method: basic theory and applications. Comput Methods Appl Mech Eng 1996;139(1–4):289–314.

[10] Remmers J, de Borst R, Needleman A. A cohesive segments method for the simulation of crack growth. Comput Mech 2003;31(1–2):69–77.